\documentclass[twocolumn]{aastex702} 

\usepackage{multirow}

\usepackage{graphics,epsf}
\usepackage[utf8]{inputenc}
\usepackage{amsmath}                
\usepackage{amsfonts}               
\usepackage{amssymb}                
\usepackage{epsfig}                 
\usepackage{graphicx}               
\usepackage{float}
\usepackage{color}
\usepackage{multirow}               

\hypersetup{
    colorlinks=true,
    linkcolor=red,   
    urlcolor=cyan}

\usepackage[colorinlistoftodos]{todonotes}

\newcommand{\s}{{~\rm s}}

\newcommand{\g}{{~\rm g}}
\newcommand{\G}{{~\rm G}}
\newcommand{\K}{{~\rm K}}

\begin{document}

\title{Comment on: ``The Preprint Evolution: The Rise of Unreviewed Drafts
on arXiv and its Implications for Astronomy''}

\author[0000-0003-0375-8987]{Noam Soker}
\affiliation{Department of Physics, Technion - Israel Institute of Technology, Haifa, 3200003, Israel; soker@technion.ac.il}
\email{soker@technion.ac.il}

\begin{abstract}
I argue against the proposal of \href{https://arxiv.org/abs/2609.35787}{arXiv:2609.35787} to divide arXiv astrophysics submissions into two streams: peer-reviewed papers and pre-review discussion drafts. I contend that such a division would threaten open scientific discussion and could lead to increasingly restrictive categories based on journal impact factor or mainstream versus non-mainstream papers. Instead, readers who want only peer-reviewed work can use AI or other tools to filter papers, while others can continue monitoring the broader literature.
I emphasize three benefits of posting papers on arXiv before journal acceptance. First, public posting lets researchers beyond the referee provide feedback, including identifying possible errors, missing references, and overlooked prior work, improving manuscripts beyond what a single referee may catch. Second, arXiv can help address weaknesses in the traditional anonymous peer-review system, such as lengthy delays, inconsistent reports, or unjustified rejection. Public posting establishes a transparent record of priority and allows the community to access results immediately. 
Finally, I propose that astrophysics journals require authors to post their papers on arXiv for at least a week before acceptance, making arXiv the central component of the astrophysical publication process.
\end{abstract}
   

\section{Introduction} 
\label{sec:intro}

 In a recent \href{https://arxiv.org/abs/2609.35787}{Short Communication}, \cite{Bedin2026} suggests splitting arXiv submissions into two streams: Level 1 (peer-reviewed science), which would host peer-reviewed papers, and Level 2 (pre-review drafts), reserved for what he terms ``discussion drafts.'' 
This, he suggests, ``. . .  would remove the unfair advantage of establishing false precedence through drafts and drastically cut the daily reading burden by allowing researchers to opt out of tracking unvetted versions.''

I think such a move would be dangerous to free scientific discussion and would start a slippery slope (Section \ref{sec:Slope}). I end with my opposite suggestion: all refereed journals force authors to post their papers on the arXiv before final acceptance (Section \ref{sec:Suggestions}).

\section{Arguments for arXiv posting} 
\label{sec:Slope}

(1) \textit{The slippery slope.} \cite{Bedin2026} argues that the $\gtrsim 100$ daily papers on the astro-ph ``places an unsustainable burden on
everyone’s time.'' However, the splitting will not reduce the number of arXiv papers, but will delay the posting of many of them. Some researchers would then argue that they do not want to waste time reading papers from low-impact-factor journals, and we will split arXiv into Q1, Q2, Q3, and Q4. Later, we will split Q1 by the journal's exact impact factor. The final step would be to split each category into mainstream and non-mainstream papers (so readers don't waste time on models and ideas that aren't mainstream). 
Therefore, splitting arXiv the way \cite{Bedin2026} suggests will harm free scientific discussion. 
The solution is much simpler: any reader can use AI to list only the accepted papers in the journal they prefer. Others, like me, will review all papers (at least their titles) to follow trends in other areas of astrophysics.  
 
(2) \textit{Improving the paper.} \cite{Bedin2026} gives a large weight to the refereeing process. But in most astronomy journals there is only one referee, who might miss many critical comments. I see two major types of comments on a new paper: (a) on giving credit to earlier papers, and (b) on errors and mistakes in the manuscript and its content. A single referee might miss some of these. Posting the paper on arXiv before acceptance allows many researchers in the field to comment to improve the paper, point out missing credit, and flag possible errors and mistakes. I post my new papers on arXiv at least two days before first submission, and have in the comment line:
\textit{It will be submitted in two days to allow for comments (including missing references).} On many occasions, I received very good comments on missing references and mistakes. 

(3) \textit{The refereeing process.} \cite{Bedin2026} refers to the posting of preprints on arXiv as ``unfair advantage of establishing false precedence through drafts.'' However, I do not think this is false precedence, but true and fair. The anonymous refereeing process is a fertile ground for power abuse. A hostile referee can simply delay their report for months (I had tens of these in my four-decade career), write a report longer than the original paper (with AI, we are getting more and more of these), or simply reject the paper with no real scientific justification, so the process begins in another journal. Each of these delays the publication in a referee journal for months to years. Posting the paper on the arXiv diminishes the power the referee has and might abuse: it lets one claim justified priority and lets the community refer to the results immediately.

\section{Mandatory paper posting on arXiv} 
\label{sec:Suggestions}

Following point 2 in Section \ref{sec:Slope} on the advantages of posting papers on the arXiv before final acceptance, I suggest that all refereed journals in astrophysics (I am not familiar with the publication culture in other fields), will demand that each paper be posted for at least a week on the arXiv before acceptance. As a referee, I usually ask authors to do so if they haven't posted their paper yet.

I had several cases where, as an arXiv reader, I pointed out to authors that they had missed earlier papers with similar claims to theirs (in some cases the authors claim they are the first) and received the reply that the paper is already accepted. With arXiv, there is no excuse for such omissions.  

In short, I suggest we boost the use of arXiv and give it the primary role in astrophysical publication. The \href{https://astro.theoj.org/about}{\textit{Open Journal of Astrophysics}} is a major step in that direction, but there are more steps for the community to take now.



\end{document}